\documentclass[%
 reprint,
superscriptaddress,
nobibnotes,
 amsmath,amssymb,
 prl,
]{revtex4-2}

\usepackage{ulem}
\usepackage{graphicx}
\usepackage{color}
\usepackage{dcolumn}
\usepackage{bm}
\usepackage{lipsum}
\usepackage{amsmath}
\usepackage{dashrule}

\begin{document}

\title{Extraordinary stiffening and reciprocal control in a DNA condensate/actin composite}
\author{Sam Wilken}
\email{swilken@uni-mainz.de}
\affiliation{Department of Chemistry, Johannes Gutenberg University Mainz, Mainz, Germany}
\author{Austin Hopkins}
\affiliation{Department of Physics, University of California, San Diego, CA 92093, USA}
\author{Kyu Hwan Choi}
\affiliation{Department of Chemical Engineering, Stanford University, Stanford, CA 94305, USA}
\author{John Berezney}
\affiliation{Department of Physics, Brandeis University, Waltham, Massachusetts 02453}
\author{Matthew E. Helgeson}
\affiliation{Department of Chemical Engineering, University of California, Santa Barbara, CA 93106, USA}
\author{M. Cristina Marchetti}
\affiliation{Department of Physics, University of California, Santa Barbara, CA 93106, USA}
\affiliation{Interdisciplinary Program in Quantitative Biosciences, University of California, Santa Barbara, CA 93106, USA}
\author{Sho Takatori}
\affiliation{Department of Chemical Engineering, Stanford University, Stanford, CA 94305, USA}
\author{Zvonimir Dogic}
\affiliation{Department of Physics, University of California, Santa Barbara, CA 93106, USA}
\affiliation{Interdisciplinary Program in Quantitative Biosciences, University of California, Santa Barbara, CA 93106, USA}
\author{Omar A. Saleh}
\email{saleh@ucsb.edu}
\affiliation{Department of Physics, University of California, Santa Barbara, CA 93106, USA}
\affiliation{Interdisciplinary Program in Quantitative Biosciences, University of California, Santa Barbara, CA 93106, USA}
\affiliation{Department of Materials, University of California, Santa Barbara, CA 93106, USA}
\date{\today}

\begin{abstract}

Soft composites, materials consisting of inclusions within a polymeric matrix, offer a broad design space with diverse mechanical properties. 
Composite stiffness is normally set by the volume fraction of filler, and classical theory predicts at most $\sim 25\%$ stiffening when inclusions occupy $10$\% of the volume. 
Here we introduce a composite of phase-separated DNA droplets that occupy $<10$\% of an entangled actin network, yet stiffen it 25-fold, reversibly.
Microscopy reveals that actin penetrates the DNA droplets, leading network elasticity to determine their size; thus the system exhibits reciprocal control between matrix and droplet properties.
We attribute the extraordinary stiffening to entrainment of actin by the droplets, which enlarges the effective inclusions until their contact network dominates the mechanics.
This shows that liquid-liquid phase separation can control not only local compositions, but also mechanics, achieving exceptional stiffening at modest filler fractions through local geometries that constrain the matrix beyond the inclusions themselves. 
\end{abstract}

\maketitle


Soft composites consisting of inclusions distributed within a polymeric or filamentous matrix are widespread in technology and biology.
The stiffness of such composites is well understood at low inclusion volume fraction, $\phi\leq 0.1$.
In this `Einstein' limit, the composite's elastic shear modulus, $G'_{comp}$, increases relative to the matrix, $G'_0$, by a stiffening factor $G'_{comp}/G'_0 = 1 + 5\phi/2$ when the inclusions are perfectly rigid~\cite{einstein_berichtigung_1911, torquato_random_2002}; at $\phi = 0.1$, this factor is only 1.25.
Larger effects are known: rigid inclusions reinforce fiber networks~\cite{islam_random_2019, shivers_criticality_2025, van_oosten_emergence_2019, yang_local_2023}, and carbon black stiffens rubber once inclusions percolate~\cite{heinrich_recent_2002, steck_multiscale_2023}, but in each case the reinforcing phase is solid, abundant, or bound to the matrix, and is built into the material irreversibly.
Liquid-liquid phase separation offers a different route, making the inclusion phase itself a controllable feature that can be created and dissolved in place by tuning a thermodynamic variable.
Liquid droplets can also stiffen composites through elastocapillary effects, though with modest stiffening factors of $\lesssim 1.5$~\cite{style_stiffening_2015,style_solidliquid_2021}.

We introduce a composite containing liquid inclusions which exceeds these behaviors, displaying extraordinary, reversible stiffening factors of up to 25 at low $\phi$.
The composite consists of phase-separated DNA droplets embedded within-- and interpenetrated by-- an entangled network of filamentous actin.
The droplets are enabled by recent advances in the sequence-programmed creation of viscoelastic liquid phases of DNA~\cite{abraham_nucleic_2024, udono_synthetic_2025}.
We combined the DNA liquid with an entangled actin network~\cite{gardel_microrheology_2003, hinner_entanglement_1998}, which serves as a matrix with orthogonal biochemistry relative to DNA, as actin and DNA do not bind one another.
Compared to the flexible-polymer networks used in other studies of droplet-embedded composites~\cite{style_stiffening_2015,style_solidliquid_2021}, actin networks display unique mechanics due to the polymer's significant bending stiffness.
This creates a persistence length, $l_p\approx 17~\mu$m, comparable to the filament length and large relative to the spacing between entanglements, altering material mechanics \cite{broedersz_modeling_2014, schmidt_chain_1989}.
Biomolecular droplets (`condensates') in semiflexible networks have been studied in biological settings~\cite{liu2023liquid,boddeker2022non}, and DNA droplets in active microtubule networks~\cite{tayar_controlling_2023,laprade_coarsening_2026}, but always as passive objects acted on by the network, rather than inclusions that modify the network's mechanics.
\begin{figure*}
    \centering
    \includegraphics[width=1\textwidth]{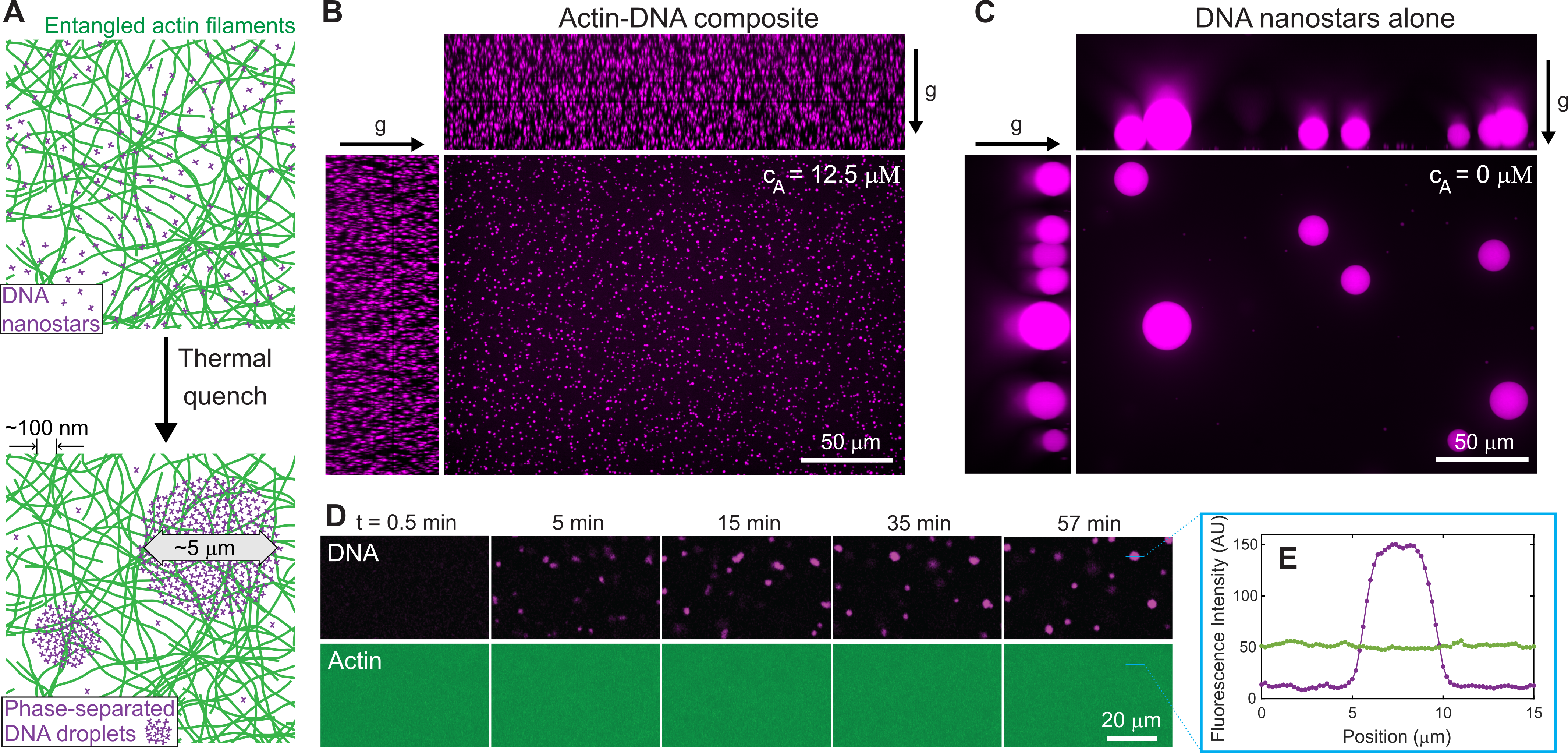}
    \caption{Design and visualization of a DNA/actin composite. {\bf A} Cartoon (not to scale) of a material comprising an entangled network of $\approx 20~\mu$m-long actin and $\approx 10$~nm branched DNA nanostars (NSs). Upon a thermal quench into the coexistence regime, NSs coagulate to form phase-separated droplets within the actin matrix. Typical length scales of droplet diameter and actin mesh size are noted. {\bf B} 3D confocal microscopy images of fluorescently labeled NSs, showing that DNA droplet growth in the presence of actin is arrested, with droplets suspended throughout the bulk. {\bf C} In contrast, droplets formed in the absence of actin exhibit uninhibited growth, and sediment to the bottom surface of the sample chamber. {\bf D} Two-color fluorescence confocal imaging of DNA and actin reveals that droplet formation does not perturb the spatial distribution of the actin network, which remains homogeneous. For all experimental conditions, the actin mesh size is smaller than the microscope's optical resolution. {\bf E} Two-channel fluorescence intensity profiles across a droplet indicate comparable actin concentrations inside and outside the droplets, consistent with actin–DNA interpenetration; this is confirmed with a correlation analysis (Supplementary Fig.~S2).} 
    \label{fig:fig1}
\end{figure*}

We find that micron-scale DNA droplets condensed around the actin network rather than expelling it, so that the DNA and actin are interpenetrated.
Network elasticity controls the droplet size, which we explain with a scaling model in which elastic hindrance to droplet Brownian motion arrests coalescence.
Quenching below the droplet melting temperature stiffened the composite up to 25-fold, at a droplet volume fraction of only $\phi \approx 0.09$.
Because the inclusions are a thermoreversible liquid phase rather than a set of permanent bonds, this large stiffness change can be switched \textit{in situ} and repeatedly, a form of control that crosslinked networks of comparable stiffness do not offer~\cite{gardel2004elastic}.
We attribute the stiffening to droplets entraining actin filaments, independent of the droplets' viscoelastic properties. Combined with the filaments' intrinsic stiffness, this extends each droplet's mechanical influence far beyond its surface, allowing these larger inclusions to percolate.
The matrix and the droplets are thus reciprocally coupled: the network controls droplet size, while droplet percolation stiffens the network.
Our results suggest mechanisms by which biological systems can effectuate large mechanical changes through control of phase separation, and offer design principles for translating those mechanisms into synthetic materials.

\begin{figure*}
    \centering
    \includegraphics[width=1\textwidth]{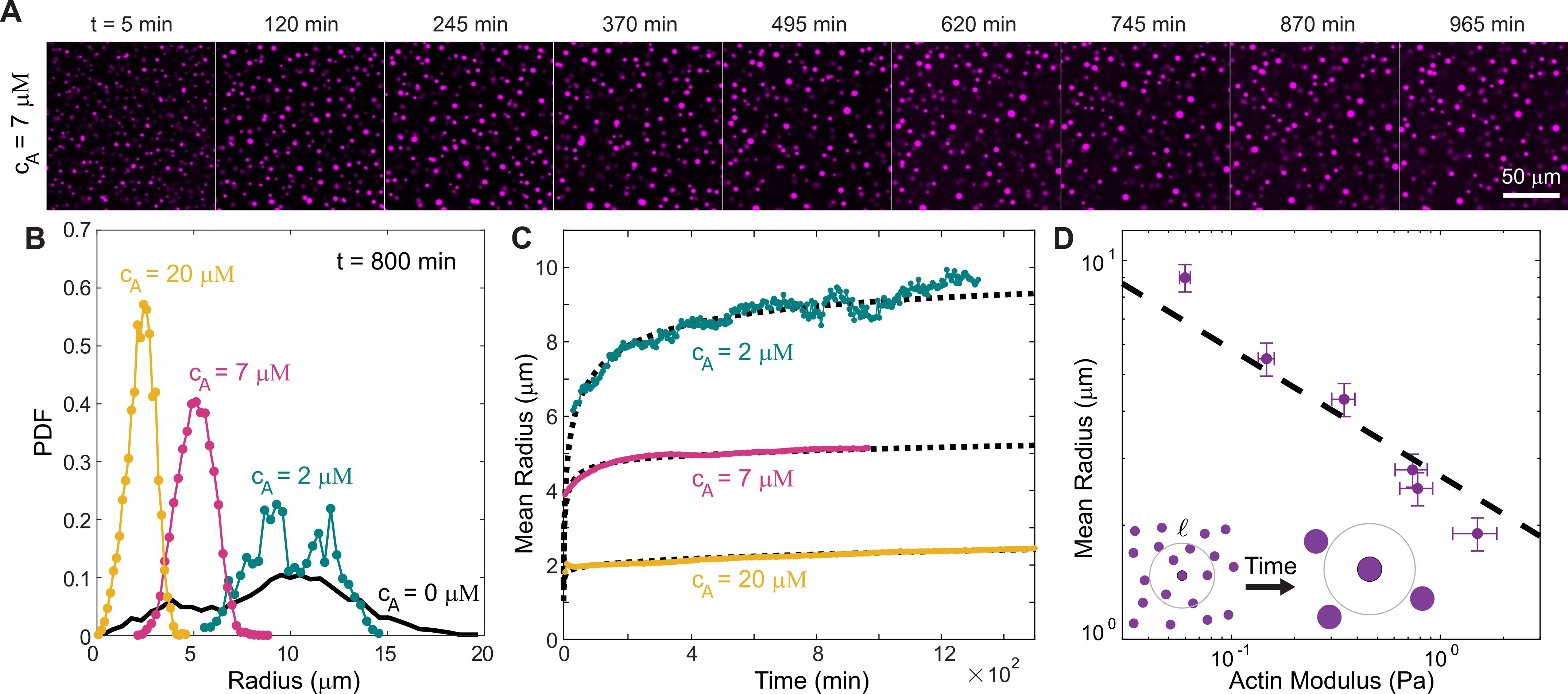}
    \caption{Actin network elasticity controls DNA droplet coarsening. {\bf A} Representative time series of confocal microscopy images showing the initial growth and subsequent dynamic arrest of fluorescently labeled DNA droplets within an unlabeled actin meshwork, with $c_A=7~\mu$M. {\bf B} Droplet size distributions $800$ min after quench, for three actin concentrations, demonstrating enhanced monodispersity and control of size, relative to samples without actin (no-actin size distributions from Ref.~\cite{wilken_spatial_2023}). {\bf C} Dynamics of the mean droplet radius versus time after quench for three actin concentrations. The dashed lines indicate an arrested-growth model comprising early-time Brownian motion and coalescence, followed by elasticity-mediated arrest of coarsening. At late times ($t>10^4$~s), and at high $c_A$, a regime of Ostwald ripening can be experimentally observed, consistent with model predictions (Supplementary Section S1 and Fig.~S4). {\bf D} Log-log plot of the mean droplet radius at $t=800$~min versus the actin network modulus, as inferred from actin concentration. The data are consistent with a model (inset cartoon) in which droplets grow via Brownian coalescence with other droplets within a distance $\ell$ determined by the effective harmonic potential imposed by the elastic network; this predicts $R\sim G'^{-1/3}$ (dashed line). All panels show data for $c_{NS} = 4~\mu$M and $T=20~^\circ$C.
    }
        \label{fig:fig2}
\end{figure*}

\subsection*{DNA droplets are interpenetrated with the actin network}
We created entangled actin networks from purified monomeric actin (concentrations $c_A = 2 - 20~\mu$M) by polymerizing the monomers at 37$~^\circ$C in standard F-buffer containing 2 mM magnesium acetate.
We fluorescently labeled the actin with phalloidin.
The filament lengths were 10-30~$\mu$m, with a mean $L=20~\mu$m (Supplementary Fig.~S1).
The final isotropic, entangled actin network constituted a weakly elastic material at moderate frequencies (see below), consistent with previous studies~\cite{gardel_microrheology_2003,hinner_entanglement_1998}.

To create inclusions, we exploited the DNA `nanostar' system. 
Nanostars (NSs) are branched DNA nanoparticles with multiple double-stranded arms ending in single-stranded sticky ends~\cite{biffi_phase_2013,abraham_nucleic_2024}.
We used fluorescently tagged 4-armed NSs and palindromic sticky ends with a sequence \textit{GGAATTCC} (full sequences in Supplementary Table S1).
Sticky-end binding then mediates homotypic NS-NS attraction, driving phase separation.
Because DNA hybridization is thermally destabilized, the condensed NS phase melts at high temperatures~\cite{biffi_phase_2013, wilken_nucleation_2024}.

We created composites by combining actin with DNA NSs at bulk concentrations $c_{NS} = 2 - 38 \mu$M. 
To ensure biomolecular homogeneity, we mixed all components at 37$^\circ$C, which is above the NS melting temperature at all concentrations used ($T_m = 27 - 31^\circ$C).
Actin polymerized for 1 hour, forming a uniform, elastic, entangled network.
Quenching the temperature below the NS melting temperature triggered phase separation (Fig.~\ref{fig:fig1}A).

After the quench, DNA droplets appeared throughout the actin network in all three dimensions, coarsening with time into roughly spherical droplets a few microns in size, and occupying total volume fractions of less than 0.1 (Fig.~\ref{fig:fig1}B).
We estimated volume fractions from 3D confocal stacks by identifying droplet centroids as local intensity maxima and assuming spherical profiles to determine droplet radii. Measured volume fractions matched lever-rule calculations from known dense, dilute, and bulk-average concentrations~\cite{wilken_nucleation_2024, abraham_nucleic_2024}.
Without actin, quench experiments were qualitatively different: the droplets coarsened to much larger sizes and sedimented to the bottom of the chamber (Fig.~\ref{fig:fig1}C).
Brownian motion and coalescence (BMC) is the dominant coarsening mechanism since the very low surface tension makes Ostwald ripening extremely slow (Supplementary Section~S1)~\cite{wilken_nucleation_2024,tateno_diffusive_2026}.
The results indicate that actin traps the NS droplets in place, suppressing both  BMC-driven coarsening and sedimentation.

We further interrogated droplet fixation using two-color fluorescence imaging of the actin network and the NS droplets.
The actin channel fluoresced uniformly prior to NS droplet formation, indicative of a mesh size less than the optical resolution limit of the microscope ($\approx 500$ nm), consistent with prior estimates ~\cite{schmidt_chain_1989}.
Surprisingly, after the thermal quench and NS droplet appearance, the actin channel remained homogeneous, even in regions occupied by droplets (Fig.~\ref{fig:fig1}D).
This homogeneity indicates that the droplets and actin meshwork interpenetrate without remodeling the actin structure.
This persists for more than 14 hours, but changes at longer times: images taken 3 days after quench showed that actin was largely excluded from the volume occupied by NS droplets (Supplementary Fig.~S2).
The initial homogeneity, and later exclusion, of actin was confirmed through an image correlation analysis (Supplementary Fig.~S2) \cite{dunn_practical_2011}.
The data show that interpenetration is the initial state of the composite: droplets nucleate and grow within a static meshwork, so filaments are enveloped rather than displaced. The equilibrium excluded state is reached over a longer timescale consistent with actin reptation (Supplementary Section S2)~\cite{kas_direct_1994,morse_viscoelasticity_1998}.

\begin{figure*}
    \centering
    \includegraphics[width=1\textwidth]{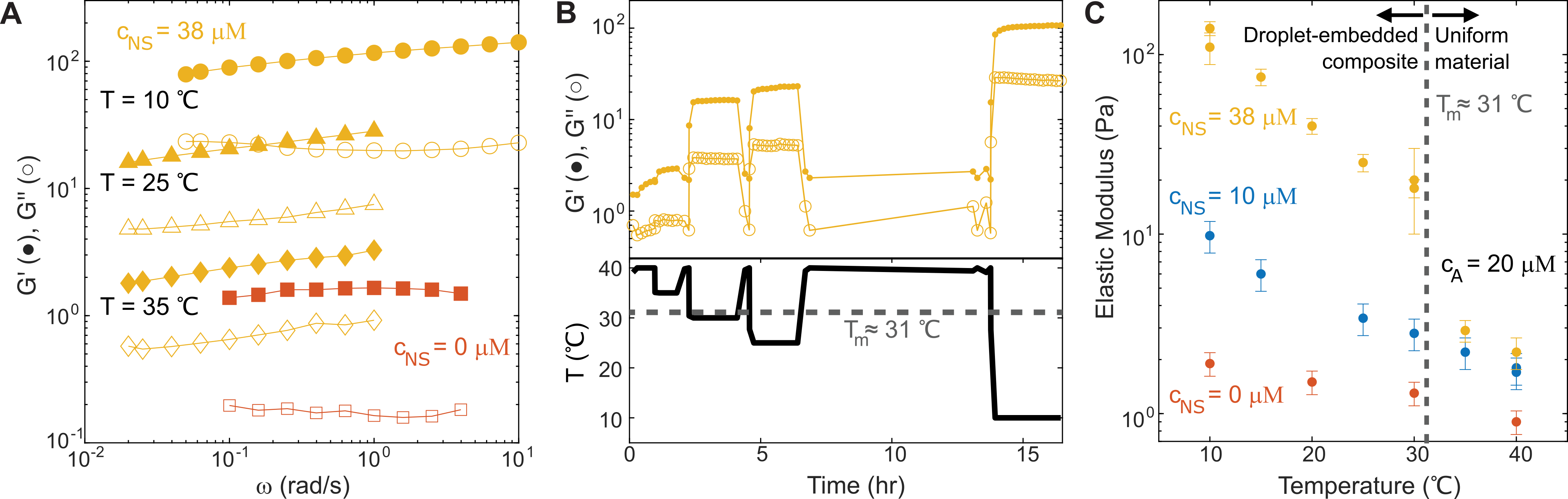}
    \caption{
Droplet-mediated stiffening. {\bf A} Bulk oscillatory shear rheology measurements of the elastic ($G'$) and loss ($G''$) moduli for actin-only networks (red), and composites of actin with nanostars (yellow; $c_{NS} = 38~\mu$M) at varying temperatures; all data used $c_A = 20~\mu$M. Across the accessible frequency range, all materials behave as soft elastic solids with only a weak frequency dependence. {\bf B} $G'$ and $G''$ measured at $\omega = 0.5$~rad/s across varying temperatures. The composite is thermoreversible: cycling the temperature across $T_m$ reproducibly switches between soft and stiff states, with the timescale of mechanical stiffening ($\approx 10$ min) comparable to that required for droplets to grow to their elastically arrested size (Fig.~\ref{fig:fig2}). {\bf C} The elastic modulus, $G'(\omega =0.5 \text{~rad/s})$, versus temperature for three different NS concentrations, in the presence of $c_A = 20~\mu$M. Actin alone shows minimal temperature dependence. The composites nearly match the actin-alone stiffness at $T>T_m$, then significantly stiffen at lower temperatures.
}
    \label{fig:fig3}
\end{figure*}

\subsection*{Actin network elasticity controls DNA droplet size}
We measured NS droplet sizes in actin from microscopy images (Fig.~\ref{fig:fig2}A; Supplementary Movie). 
The size of embedded droplets was only weakly dependent on $c_{NS}$ at constant $c_A$ (Supplementary Fig.~S3). 
Conversely, at constant $c_{NS}=4~\mu$M, droplet size decreased with $c_A$, and the droplets were more monodisperse compared to when actin was absent (Fig.~\ref{fig:fig2}B). 
BMC-driven thermodynamic fluctuations tend to increase droplet polydispersity; thus, the monodispersity observed with actin results from reduced droplet mobility.

Coarsening measurements showed that droplets initially grew quickly, in quantitative agreement with BMC models (Supplementary Section S1). 
After a few hours, the mean size reached a plateau that decreased with increasing $c_A$ (Fig.~\ref{fig:fig2}C).
At higher $c_A$, where droplet sizes were smaller, the mean droplet size increased at long times ($\approx 10^4$~s; Supplementary Figure S4), consistent with larger Laplace pressures causing a slow Ostwald process. This multi-regime coarsening is supported by a model combining BMC and Ostwald mechanisms~\cite{lifshitz_kinetics_1961}, which fit the data well (Supplementary Section S1 and Fig.~S4).

We account for the dependence of droplet size on actin concentration at intermediate times ($\approx 800$ min) by assuming that droplets undergo early-time Brownian motion that is hindered by the surrounding actin network.
We assume that the entangled network imposes a harmonic elastic constraint on thermally-driven droplet motion, such that the maximum distance, $\ell$, reached by a droplet of radius $R$ embedded within a network of modulus $G'$ is $\ell \approx R+ \sqrt{k_BT/G'R}$.
Droplets can contact others within $\ell$, and grow through coalescence, which proceeds until there are no other droplets within $\ell$.
The final length scale, $\ell_f$, thus also characterizes the inter-droplet spacing, so the volume fraction is $\phi \approx R_f^3/\ell_f^3$.
As $\phi$ is a constant, unchanged by coalescence, we have $R_f \approx \phi^{1/3} \ell_f$, giving $R_f \sim G'^{-1/3}$.
This scaling matches experimental measurements of mean droplet radius vs.~actin network elasticity (Fig.~\ref{fig:fig2}D), where we infer elasticity from the established scaling $G' \sim c_A^{7/5}$ \cite{hinner_entanglement_1998,gardel_microrheology_2003,broedersz_modeling_2014}.
In other systems, matrix elasticity can thermodynamically select a length scale that determines the domain size of elastic microphase separation~\cite{fernandez2024elastic}, can reshape the phase behavior of embedded protein condensates~\cite{fritzen2026programmable}, and can even suppress phase separation outright~\cite{rosowski2020elastic}.
Here, by contrast, elasticity arrests coalescence kinetically.

\begin{figure*}
    \centering
    \includegraphics[width=\textwidth]{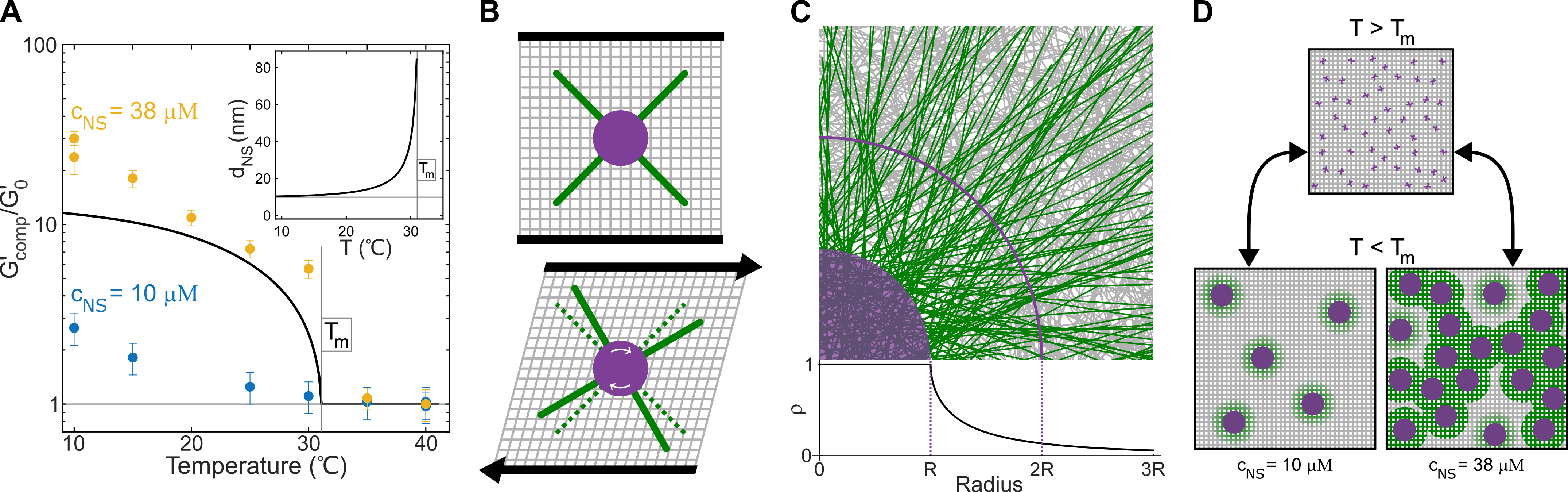}
  \caption{
Composite stiffening factor and model.
{\bf A} Stiffening factor, $G'_{comp}/G'_0$ at $\omega=0.5$~rad/s, vs.~temperature, where $G'_0$ is the droplet-free modulus inferred from the actin-only data and the modest shift caused by dilute NSs (Supplementary Fig.~S5).
The line is the confinement-tube prediction of stiffening for nanostar concentration $c_{NS} = 38~\mu$M, evaluated using the $T$-dependent NS tube size $d_{NS}(T)$ plotted in the inset.
{\bf B} Bulk shear rotates a droplet (purple) threaded by actin filaments (green), dragging them through the non-penetrating matrix (gray) and extending the droplet's mechanical influence beyond its surface.
{\bf C} Top: Schematic of the high density of penetrating filaments next to a droplet, shown as one quadrant of a 2D construction in which lines of length $10R$ are placed at random positions and orientations about a disc of radius $R$, colored green or gray according to whether they intersect the disc.
Bottom: Matrix composition $\rho$ on the same radial axis, computed for the experimentally relevant 3D geometry; $\rho=1$ ($0$) denotes a matrix composed entirely of penetrating (non-penetrating) filaments.
The arc at radius $2R$ marks the ansatz, $\lambda = R$ of the distance of entrainment beyond the droplet surface.
{\bf D} Cooling through $T_m$ reversibly converts a homogeneous NS-actin solution (top) into droplets (purple) enlarged by entrained actin (green halo). At high $c_{NS}$ these enlarged inclusions percolate into a connected, rigid subphase.}
    \label{fig:fig4}
\end{figure*}

\subsection*{Actin/DNA droplet systems display extraordinary stiffening}
We next examined how droplets affect the composite's mechanical properties. 
Solutions of monomeric actin and NSs were prepared at 37°C and transferred to a rheometer at the same temperature, allowing actin to polymerize {\it in situ} for 1 hour. 
Subsequently, the temperature was reduced to induce NS phase separation, and the sample was equilibrated for 1 hour.
Then, small amplitude oscillatory rheological measurements were taken at various frequencies, $\omega$ (Fig.~\ref{fig:fig3}A).
After each test, the temperature increased to $40^\circ$C to melt the NSs, as confirmed by rheology (Fig.~\ref{fig:fig3}B), then cooled again for the next measurement. 
This protocol ensured near-equilibrium structures are measured without influence from previous droplet arrangements or tests.

Oscillatory rheological measurements showed little dynamic effect: over the accessible frequency range, both actin-only and actin/NS composites acted as elastic materials ($G'\gg G''$), with moduli that weakly increase with $\omega$ (Fig.~\ref{fig:fig3}A). 
There was thus no major dissipative relaxation over the measured timescales. For actin, this is likely because reptation is slow for the lengths utilized (Supplementary Section~S2) ~\cite{kas_direct_1994,morse_viscoelasticity_1998}. 
NS liquids can show relaxations in the measured frequency regime, either from bond dynamics~\cite{biffi_phase_2013, conrad_increasing_2019} or droplet capillary relaxation~\cite{jeon_salt-dependent_2018}. 
Their absence here indicates NS droplet dynamics aren't a major factor in the system's mechanics.

The composite had a higher plateau modulus and greater temperature dependence than actin alone (Fig.~\ref{fig:fig3}C).
Without NSs, and at $c_A = 20~\mu$M, the networks had a modulus $G' \approx 1$~Pa at high temperature, similar to prior work~\cite{gardel_microrheology_2003,hinner_entanglement_1998}; this modestly increased, by twofold, at low temperature.
With NSs and at high temperatures ($T>T_m$), the composite modulus $G'_{comp}$ doubled compared to the NS-free case (Fig.~\ref{fig:fig3}C), likely due to the dissolved, homogeneous NSs.
In contrast, as the temperature was lowered, $G'_{comp}$ strongly increased, in a fully thermo-reversible fashion (Fig.~\ref{fig:fig3}B, C).
At $c_{NS} = 38~\mu$M, $G'_{comp}$ increased discretely as $T$ dropped below $T_m$, indicating droplet formation's role, and kept rising at lower temperatures.
To estimate an inclusion stiffening factor, we define an apparent modulus, $G'_0(c_{NS},T)$, based on the actin-only $G'(T)$ but including dissolved NS effects (Supplementary Fig.~S5). 
The stiffening factor, $G'_{comp}/G'_0$, shows that droplet formation alone increases stiffness by about 25-fold at the lowest temperatures (Fig.~\ref{fig:fig4}A).

\subsection*{Tight actin confinement causes NS droplets to act as rigid inclusions}
We sought a mechanistic basis for the large stiffening effects induced by DNA droplets.
The mechanics of other droplet/matrix composites can be explained by elastocapillary effects~\cite{style_stiffening_2015, style_solidliquid_2021}.
However, such effects are negligible here, both because the droplet volume fractions are very small ($<0.1$), and because of the small value of the NS droplet surface tension, $\sigma \approx 10^{-6}$ N/m~\cite{jeon_salt-dependent_2018}, giving $\sigma/R \lesssim 1$~Pa $\ll G'_{comp} \approx 10-25$~Pa.
We considered whether NSs could non-specifically bind and crosslink the actin network, as crosslinking proteins can raise the network modulus by orders of magnitude~\cite{gardel2004elastic,lieleg_structure_2010, broedersz_modeling_2014}.
But this would also lead to strain-stiffening~\cite{semmrich_nonlinear_2008}, and nonlinear rheology measurements showed that both composites and actin-only networks purely strain-softened (Supplementary Fig.~S6), indicating there is no appreciable crosslinking in either system.

Instead, we attribute the stiffening to the tight confinement of actin filaments within the droplets and to the associated effects on the actin meshwork immediately adjacent to them.
Confining a filament induces an entropic cost by reducing the range of its thermal bending fluctuations.
In entangled actin-only networks, confinement of each filament by the surrounding meshwork determines the network stiffness, with established models predicting the modulus varies as $G' \sim d^{-2/3}$, where $d$ is the diameter of the effective confining tube created by the actin mesh (with $d\approx 140$~nm at $c_A = 20~\mu$M)~\cite{broedersz_modeling_2014}.

Prior work found that NS droplets present an effective confining diameter, $d_{NS} \approx 10$~nm, to embedded polymers~\cite{nguyen_length-dependence_2019}.
$d_{NS}$ was interpreted as the typical scale of cavities created in the NS liquid by thermal fluctuations~\cite{nguyen_length-dependence_2019}, and thus can be taken as the equilibrium diameter of a tube that confines embedded actin.
A confinement-tube picture is directly relevant at lower temperatures ($T\leq 20^\circ$C), where the viscoelastic behavior of the NS liquid dictates it will be solid-like on the timescales of rheological measurement (Supplementary Table S2) \cite{conrad_increasing_2019}. While the droplets are liquid-like on these timescales at higher temperatures ($T=25,30^\circ$C), the similarities in the dynamic rheology and modulus indicate a similar mechanism still holds, though might be due to a kinetic clamping of the actin by the high viscosity of the NS liquid.
The small value of $d_{NS}$ indicates that actin threaded through a droplet is tightly confined, on a scale barely larger than the $d_A = 7$ nm diameter of the filament itself~\cite{hanson_structure_1963}.
That $d_{NS} \ll d$ suggests a large confinement penalty for filaments inside the droplet relative to the bulk, which, along with demixing, is likely related to the tendency of the actin to exit the droplet after long times (Supplementary Fig.~S2).

The tight confinement presented by the NS droplet interior will substantially stiffen the contained actin meshwork, causing the droplet/actin inclusion to act as a rigid element within a softer, actin-only matrix.
However,  the small droplet volume fractions ($\phi_{NS} = 0.025$ and $0.09$ for, respectively, $c_{NS} = 10~\mu$M and $38~\mu$M) fall within the Einstein limit discussed above~\cite{torquato_random_2002}, and so even perfectly rigid droplets cannot account for the observed stiffening.

\subsection*{Droplets entrain nearby actin, enlarging inclusions}
We suggest that confining actin within NS droplets, together with actin's intrinsic stiffness (persistence), leads to an effective inclusion radius that is up to twice that of the NS droplet alone.
We attribute this enlargement to the tendency of a stiff inclusion to rotate when the surrounding soft matrix is sheared: because simple shear is a combination of a pure strain and a rigid rotation, and the former is not possible for a stiff inclusion, it must rotate~\cite{landau_course_1959,puljiz_forces_2017}.
Actin filaments within the droplet will be clamped by tight confinement and entrained with the rotating inclusion.
This will cause the filaments' external segments to sweep out an angular trajectory, creating mechanical effects at distances far from the droplet surface (Fig.~\ref{fig:fig4}B).
The elasticity of the surrounding meshwork impedes this motion, so the entrainment effect is screened with increasing distance from the droplet surface.
The increased size of the effective inclusion can then be defined by a mechanical screening length, $\lambda$.

While $\lambda$ could relate to the behavior of a single filament (Supplementary Section~S3 and Fig.~S7)~\cite{landau_course_1959,broedersz_modeling_2014}, the dominant contribution likely arises from interactions between the large number of droplet-embedded filaments.
Estimates using $c_A=20~\mu$M and $R= 2~\mu$m indicate each droplet contains $\approx 500$ filaments (Supplementary Section S4). 
Thus, the droplet-adjacent matrix will mainly consist of droplet-embedded filaments (Fig.~\ref{fig:fig4}C) that are entrained with the droplet because each filament will tend to move with its neighbors.
We estimate the screening length by quantifying this cooperative effect, particularly by assuming the matrix remains entrained until its fraction of droplet-intersecting filaments becomes negligible.
In the limit $L,l_p \gg R$, geometry dictates that the fraction of droplet-intersecting filaments at radial distance $r$ from the droplet surface is $\rho(r) = 1-\sqrt{1-(R/(R+r))^2}$ (Fig.~\ref{fig:fig4}C); this decay indicates that $\lambda$ scales with the droplet radius.

Extrapolating from the scaling relation, we make the ansatz that $\lambda = R$, and find that this estimate consistently describes the data.
In this view, the effective inclusion radius is $R+\lambda = 2R$, and the effective volume fraction is $\phi_{eff}\equiv \phi_{NS}((R+\lambda)/R)^3 = 8\phi_{NS}$
For $c_{NS} = 10~\mu$M, with droplet volume fraction $\phi_{NS}\approx 0.025$, this means $\phi_{eff} = 0.2$. 
At such a concentration, the inclusions remain discrete, and composite elasticity can be estimated from effective-medium models of inclusion mechanics \cite{torquato_random_2002}. 
These models predict a stiffening factor of 1.8 at $\phi = 0.2$ (Supplementary Section S5), consistent with the maximal measured stiffening factor of $G'_{comp}/G'_0 =2.2 \,\pm\ 0.3$ (averaged from lowest two temperatures).

The $\lambda = R$ ansatz also accounts for the $c_{NS}=38~\mu$M behavior, where  $\phi_{NS} = 0.09$, so 8-fold enlargement leads to $\phi_{eff} \approx 0.7$.
This exceeds the percolation threshold for inclusions connected by overlapping shells: 
In our system, shells interpenetrate, but droplet cores do not, and the percolation threshold for such geometries has been calculated to be $\phi_c \approx 0.3\pm0.1$ over a broad range of core-to-shell ratios~\cite{bug_interactions_1985}.
Percolation of entrainment shells indicates inclusions are connected by shared actin filaments which transmit torque (Supplementary Fig.~S8); thus, the network will be rigid~\cite{feng_percolation_1984}. 
Further, recent work on the mechanics of a hydrogel with solid fillers attributed stiffening to percolation of enlarged inclusions \cite{dellatolas_local_2023}, indirectly supporting the mechanism proposed here. 
We conclude that a rigidity percolation transition provides a compelling rationale for the large stiffening of the $c_{NS}=38~\mu$M system.

Percolation reasonably explains the discrete stiffness change at $T=T_m$ but not its magnitude or temperature dependence at $T<T_m$ (Fig.~\ref{fig:fig4}A). 
The percolated cluster's stiffness saturates at that of the confined actin, which depends on the degree of confinement, $d_{NS}$.
As $d_{NS}$ describes the size of cavities created by thermal fluctuations \cite{nguyen_length-dependence_2019}, it is likely to scale with the thermal capillary length, $d_{NS} \sim \sqrt{k_BT/\sigma}$.
We can then estimate $d_{NS}$ vs. $T$ using estimates of the temperature dependence of $\sigma$ \cite{wilken_nucleation_2024} (inset, Fig.~\ref{fig:fig4}A), and, in turn, approximate the temperature-dependent stiffening factor from the scaling picture of confinement, accounting for actin diameter, as $G'_{comp}/G_0 = \left[(d-d_A)/(d_{NS}(T)-d_A)\right]^{2/3}$. 
This prediction roughly accounts for the magnitude and thermal behavior of the $c_{NS} = 38~\mu$M data (Fig.~\ref{fig:fig4}A).


\subsection*{Discussion}
We present a study of the structure and mechanics of a composite consisting of an actin network modulated by phase separation of a biomolecular condensate.
This material displays unique emergent behaviors, including matrix-controlled droplet size and monodispersity, and 25-fold stiffening from droplet formation, despite droplets occupying under 10\% of the volume.
The system shows reciprocal mechanical control: network elasticity governs droplet size, while droplet percolation controls network stiffness.
We emphasize that this coupling is reciprocal, not feedback, as it consists of two independent, one-way influences acting in opposite directions.
It is unclear if network-controlled droplet size influences the observed stiffening, rather than the droplet volume fraction.
Our interpretation suggests that effective-inclusion enlargement depends on the relative droplet size, not the absolute size, decoupling the network's control of size from its role in stiffening.
However, our interpretation is of an isolated inclusion. 
When entrainment shells overlap, shared filaments between droplets cannot co-rotate, causing rotational frustration and stress, which a single-inclusion view omits~\cite{puljiz_forces_2017}.
In such multi-inclusion mechanisms, droplet size and spacing will interact with filament contour and persistence lengths to influence mechanics, creating feedback. Closing this loop is a goal for future work.

We have further related these emergent behaviors to the specific behaviors of the constituent biomolecules, providing physical insight into the microscopic mechanisms in play.
Notably, due to the low surface tension of DNA droplets, elastocapillarity is irrelevant~\cite{style_liquid-liquid_2018,qiang_nonlocal_2024}; we instead present a mechanism where coalescence-driven coarsening, hindered by matrix elasticity, explains the variation of droplet size. 
Further, because the actin network threads through the droplets, the inclusion and matrix are coupled along nanostar-embedded filaments.
This interpenetration leads to the stiffening effect: each droplet confines the actin passing through it, and these captured filaments determine an inclusion's mechanical size.
This geometry is distinct from condensates that adhere non-specifically to filaments and locally reorganize them~\cite{boddeker2022non}, which we do not observe.
Confinement enables persistence of the actin filaments to transduce shear-induced rotation of the inclusions into a substantially enlarged effective inclusion size. 
The resulting effective volume fraction ($\phi_{eff} \approx 0.7$) is well above the percolation threshold of such interpenetrating inclusions ($\phi_c \approx 0.3$), rationalizing the large observed stiffening.
This stiffening remains finite, not diverging as would occur for rigid inclusions, because the fully percolated contact network is only as stiff as the confined actin matrix within each droplet.

Our results have implications for composite mechanics in biology, where condensates and filamentous polymers are widespread features of the cytoskeleton~\cite{kato_wetting-mediated_2026, setru_hydrodynamic_2021}.
They suggest that biochemical control of condensate formation can adaptively transmit large mechanical changes to the surrounding network.
More generally, our work continues a recent focus on the reciprocal effects between biological condensate formation and the surrounding medium \cite{liu2023liquid,boddeker2022non,kothari_crucial_2023}.
Further, the system's unique mechanics enable materials behaviors of interest. 
Our work presents an alternative, surfactant-free pathway to controlling droplet monodispersity. 
The stiffening is also thermally switchable: composite modulus reproducibly toggles between soft and stiff states as temperature is cycled across the droplet melting point, with switching kinetics ($\sim$10 min) matching the timescale of elastically arrested droplet growth. 
Because phase separation is an intrinsically nonlinear, stimulus-responsive transition, this points toward leveraging these effects to create responsive, switchable materials~\cite{jaspers2017nonlinear,xue2025hydrogels}.


\subsection*{Online Methods}
\subsubsection*{Actin preparation}
Rabbit skeletal muscle actin is purified from muscle acetone powder (Pel-Freez, catalog no: 41995-2, Lot 16743) following the methods in \cite{Spudich1971, MacLean-Fletcher1980}.
No rabbits or other animals are directly involved in this study.
Actin is stored as depolymerized globular actin (G-actin) at $-80$\textdegree C in G-buffer with 6\% sucrose until use.
Filamentous actin buffer (F-buffer) consists of 50 mM Tris (pH 7.5), 2 mM magnesium acetate, 0.5 mM adenosine triphosphate (ATP), 0.2 mM calcium chloride, 25 mM potassium acetate, and 1 mM dithiothreitol (DTT). DTT is added to all buffers immediately before use to preserve its reactivity.
Globular actin buffer (G-buffer) consists of 2 mM Tris (pH 8.0), 0.2 mM calcium chloride, 0.5 mM DTT, 1 mM sodium azide, and 0.2 mM ATP.
Phalloidin, used to fluorescently visualize the actin, was added at 1:1 stoichiometry with actin monomers, with a 1:100 ratio of Atto488-labeled to unlabeled phalloidin.
Actin monomers were mixed with F-buffer, and immediately loaded into a sample chamber using a trimmed pipette tip to limit shear-induced scission of actin. The mixture was allowed to incubate for 1 hour to create the entangled actin network.

\subsubsection*{Nanostar preparation}
Unlabeled constituent oligonucleotides were synthesized and desalted by Integrated DNA Technologies (IDT), resuspended in 10 mM Tris-HCl (pH 7.4), and filtered/reconcentrated using 100 kDa and 3 kDa centrifugal filters. 10\% of nanostars were dyed with Cy5 fluorophore, achieved by substituting 10\% of one arm-forming strand with a 3'-Cy5-labeled version; this labeled strand was synthesized and HPLC-purified by IDT and resuspended in 10 mM Tris-HCl. The four constituent strands were combined in equimolar ratio in 10 mM Tris-HCl, 250 mM NaCl and annealed by incubating at 95$^{\circ}$C for 10 min, then cooling to 4$^{\circ}$C at $-0.5^{\circ}$C/min. Annealed nanostar stocks were stored at $-20^{\circ}$C until use.

\subsubsection*{Microscopy}
Two-color fluorescence imaging of actin/nanostar composites was performed on a Leica SP8 confocal microscope using a 40$\times$ water-immersion objective. Atto488-phalloidin (actin) and Cy5-nanostar (NS) channels were acquired sequentially, line-by-line, to eliminate cross-talk.
 
Because a droplet's finite size is not much larger than the axial (z) resolution of the confocal system, out-of-plane fluorescence can bleed into an otherwise non-fluorescent region and limit the apparent completeness of any true fluorescence exclusion; we performed a control measurement to quantify this imaging artifact. Undyed polystyrene colloids of comparable size to NS droplets were sedimented in a solution of free Atto488-phalloidin at a fluorophore concentration matching that used in the actin-exclusion experiments ($\approx0.1~\mu$M) and imaged under the same conditions (Supplementary Fig.~S9). This control establishes the imaging-only baseline for apparent fluorescence exclusion, against which the genuine, long-timescale actin exclusion from NS droplets (Supplementary Fig.~S2) can be compared.
 
Droplet size and growth-kinetics measurements (Fig.~\ref{fig:fig2}) were performed using a Nikon Crest V3 spinning-disk confocal microscope with a 20$\times$ air objective.

\subsubsection*{Rheology}
Bulk oscillatory shear rheology was performed on a TA Instruments AR-G2 stress-controlled rheometer in a parallel-plate geometry. The top plate was a 22 mm diameter glass disc affixed to the rheometer tool with double-sided tape; the bottom plate was a silane-treated glass microscope slide taped to a Peltier-controlled bottom plate for temperature control. Both plates were silane-treated to encourage actin surface anchoring~\cite{sanchez_engineering_2013}. The sample was surrounded with mineral oil containing 2\% Span 80 to prevent sample evaporation and actin adhesion to the oil-water interface. All rheology measurements were performed with a gap height of 300 $\mu$m (sample volume $\approx$120 $\mu$L). All frequency sweep measurements were performed at 5\% strain, in the linear elastic regime for both samples with and without droplets (Supplementary Fig.~S6).

\subsubsection*{Statistics and error bars}
Error bars in Fig.~\ref{fig:fig2}D reflect two independent sources of uncertainty: the vertical (mean droplet radius) error is the fitting uncertainty of the elastically-arrested growth model (Supplementary Eq.~S4) to the radius-vs-time data, while the horizontal (actin network modulus) error is propagated from an estimated 10\% pipetting uncertainty in the actin monomer concentration, $c_A$, via the scaling $G' \sim c_A^{7/5}$. Each actin concentration in Fig.~\ref{fig:fig2}D corresponds to a single sample.

Error bars in Fig.~\ref{fig:fig3}C and Fig.~\ref{fig:fig4}A (which uses the rheology data shown in Fig.~\ref{fig:fig3}C) represent the standard error of the mean of bulk rheology measurements. Data at $c_{NS} = 10$ and $38~\mu$M each derive from repeated quenches after resetting to a homogeneous state in two independent samples; data at $c_{NS} = 0$ µM (actin alone) derive from a single sample, with three additional actin-only samples (not shown) falling within the plotted error bars.

\subsection*{Author contributions}
O.A.S., Z.D., and S.T. conceived the project and, with S.W., designed experiments. S.W. carried out experiments, with assistance from J.B. and K.H.C. M.H. assisted with rheology. A.H. and M.C.M developed scaling arguments. S.W. and O.A.S. analyzed and interpreted data. The paper was written by S.W. and O.A.S., with assistance from all co-authors.

\subsection*{Acknowledgements}
This work was supported by the MRSEC Program of the National Science Foundation under Award No. DMR 2308708. We thank Nicholas Cuccia, Sattvic Ray, and Anukta Datta for assistance with the rheology and actin preparation.

\subsection*{Data availability}
The data generated in this study, including the raw data and code underlying the figures and results reported herein, will be deposited in the Dryad Digital Repository and made publicly available upon publication through a DOI that will be included here.

\end{document}